\magnification1200

%\rightline{KCL-MTH-14-05}
%\rightline{hep-th/yymmnnn}

\vskip 2cm
\centerline
{\bf  $E_{11}$ must be a symmetry of strings and branes}
\vskip 1cm
\centerline{ Alexander G. Tumanov and Peter West}
\centerline{Department of Mathematics}
\centerline{King's College, London WC2R 2LS, UK}
\vskip 2cm
%\centerline{and}
%\vskip 0.5cm
%\centerline{,}
%\centerline {,}
%\centerline{}
\leftline{\sl Abstract}
We construct the non-linear realisation of the semi-direct product of $E_{11}$ and its vector representation in five and eleven dimensions and find the dynamical equations it predicts at low levels. Restricting these results to contain only the usual fields of supergravity and the generalised  space-time to be the usual space-time we find the equations of motion of the five and eleven dimensional maximal supergravity theories. Since this non-linear realisation contains effects that are  beyond the supergravity approximation and are thought to be present in an underlying theory we conclude  that the low energy effective action of string and branes must possess an $E_{11}$ symmetry. 
\vskip2cm
\noindent

\vskip .5cm

\vfill
\eject
{\bf 1 Introduction}
\medskip
It has been conjectured that the low energy effective action for strings and branes is the non-linear realisation of the semi-direct product of $E_{11}$ and its vector ($l_1$) representation,  denoted $E_{11}\otimes _s l_1$ [1,2].  This theory has an infinite number of fields, associated with $E_{11}$,  which live on a generalised spacetime, associated with the vector representation $l_1$. The fields obey equations of motion that are a consequence of the non-linear realisation.  
\par
For many years,  it was believed that strings by themselves could constitute a complete consistent theory whose precise form was yet to be found. However, while the perturbative quantum properties of string theory are very  well understood, the situation is much less clear for  their non-perturbative properties. Although supergravity theories in ten dimensions were not believed to be  consistent theories of quantum gravity they did provide a low energy effective action for strings which included their non-perturbative effects. It became clear from these supergravity theories that the underlying theory must contains branes as well as strings. Unfortunately,  it is not understood how  to quantise a single brane and not much is known about how branes scatter. Thus it became clear that very little is known about the underlying consistent quantum theory that contains strings and branes. 
\par
For simplicity we   will now restrict our discussion to maximally supersymmetric theories.   Although, as we have just indicated,  supergravity theories have provide a very useful starting point for discussions of an underlying theory there are quite a number of maximal supergravity theories. These is  one in eleven dimensions, two in ten dimensions and one in each dimension below ten. In addition  there also a substantial number of so called gauged supergravity theories which are obtained by adding a cosmological constant to a massless  maximal supergravity theory in a given dimension and then seeing what theories are possible whilst preserving all the supersymmetries [3]. 
\par
Although  there are quite a few  maximal supergravity theories, and so string theories,   some of them  are related to each other by duality transformation; such as the  T duality between the IIA and IIB supergravity theories in ten dimensions and the relation between the IIA supergravity theory and the eleven supergravity  theory. 
Such relations were summarised in what became known as M theory, but  M theory is not so much a unified theory but  more a set of observations between different theories. Furthermore,  M theory does not include relations between  the different gauge supergravities,  or relate them to  the massless supergravities  in a systematic way. 
\par
The situation with $E_{11}$ is different. The different theories emerge by 
decomposing $E_{11}$  into different subgroups, in particular,  the theory in $D$ dimensions arises by considering the decomposition into $GL(D) \otimes E_{11-d}$,  except in ten dimensions where there are two possible decompositions corresponding to the IIA and IIB theories [4,5,6,7]. The fields one finds at low levels in these different  decompositions   are just the fields of the corresponding supergravity theories.   However, at the next level  level one  finds next to spacefilling forms (fields with $D-1$ totally anti-symmetric indices) that lead to all the different gauged supergravities [6,8]. As a result all the maximal supergravity theories, including the gauged supergravities,  are contained in a single theory, namely the non-linear realisation of $E_{11}\otimes l_1$. Some of the fields that arise  at higher levels in $E_{11}$  are known to be need. One example is the ten forms predicted [9] in ten dimensions for the IIA and IIB theories which were confirmed to be present by showing that they are the unique such forms that are admitted by the supersymmetry algebras with which these theories are usually formulated [10]. 
\par
There is also very good evidence that all the brane charges are contained in the vector representation [2,11,12,13]. However, the situation is less clear for the physical relevance of the infinite number of coordinates beyond those of the usual spacetime that arise in the $E_{11}\otimes_s l_1$ non-linear realisation. Nonetheless, all  gauged maximal supergravities in five dimensions were  constructed by taking the fields to depend on some of the higher level generalised coordinates [7].
\par
Partial results on the dynamics encoded in the $E_{11}\otimes_s l_1$ non-linear realisation can be found in many of the early $E_{11}$ papers. However, these papers often used only the usual coordinates of spacetime and also only  the Lorentz part of the $I_c(E_{11})$ local symmetry, as a result the full power of the symmetries of the non-linear realisation was not exploited.  A more systematic approach was used  to constructing  the equations of motion of the  $E_{11}\otimes_s l_1$ non-linear realisation 
 in eleven [15]  and four [16] dimensions by including  both the higher level generalised coordinates and local symmetries in $I_c(E_{11})$. The equations of motion of the form fields were derived,  but these papers did not complete the derivation of the gravity equations and as a result the complete contact with the supergravity theories was not made. A more detailed account of  previous papers which contained partial derivations of the equations of motion of the non-linear realisation $E_{11}\otimes_s l_1$ can be found in [17]
\par
In this paper we will rectify the just mentioned  defect. We will, as before,  derive the equations of motion for the form and scalar fields,  this time in five dimensions, but we will also find the gravity equation at low level. The result is that if, at the end of  the derivation, we systematically truncate to  keep only the usual fields of supergravity and derivatives with respect to the usual coordinates of spacetime then the equations of motion are those of five dimensional maximal supergravity. In doing this  we will chose the values of  two constants which are not determined as a result of the limited extent to  which  the calculation has been carried out in this paper. These constants will,  however, be fixed by taking subsequent variations of the equations and this will be given in a future publication. 
\par
In section four we  will also evaluate the $E_{11}\otimes_s l_1$ non-linear realisation in eleven dimensions. When restricted in the same way we find the equations of motion of eleven dimensional supergravity. However, in this case we will vary the gravity equation under the symmetries of the non-linear realisation, at the linearised level,  to find the three form equation of motion. In so doing so we will  fix the remaining undetermined constant. Thus we precisely recover the bosonic  sector of eleven dimensional supergravity. As a result we have in effect shown the $E_{11}$ conjecture [1,2];  namely that the $E_{11}\otimes_s l_1$ non-linear realisation provides a unified  low energy effective theory of strings and branes.  
\par
We begin by very briefly recalling  the main features of the non-linear realisation of $E_{11}\otimes_sl_1$ which is constructed from the group element $g\in E_{11}\otimes_sl_1$ that  can be written as 
$$
g=g_lg_E 
\eqno(1.1)$$
In this equation  $g_E$ is a group element of $E_{11}$ and so can be written in the form 
$g_E=e^{A_{\underline \alpha} R^{\underline \alpha}}$ where the $R^{\underline \alpha}$ are the generators of $E_{11}$ and $A_{\underline\alpha}$ are the fields in the non-realisation. The group element   $g_l$ is formed from the generators of the vector ($l_1$) representation and so has the form $e^{z^A L_A} $ where $z^A$ are the coordinates of the generalised space-time. The fields $A_{\underline\alpha}$ depend on the coordinates $z^A$.  The $E_{11}\otimes_s l_1$ algebra and the explicit form of these group elements  can be found in earlier papers on $E_{11}$,  for example in dimensions eleven [15], five [7] and four [16]. 
The non-linear realisation is, by definition, invariant under the transformations 
$$
g\to g_0 g, \ \ \ g_0\in E_{11}\otimes _s l_1,\ \ {\rm as \  well \  as} \ \ \ g\to gh, \ \ \ h\in
I_c(E_{11})
\eqno(1.2)$$
The group element $g_0\in E_{11}$ is a rigid transformation, that is, it is  a constant. The group element $h$ belongs to the Cartan involution invariant subalgebra of $E_{11}$, denoted $I_c(E_{11})$; it is a local
transformation meaning that  it depends on the generalised space-time. 
The action of the Cartan involution can be taken to be 
$I_c(R^{\underline \alpha}) = - R^{-\underline \alpha} $ 
for any root $\alpha$ and so the Cartan involution invariant subalgebra is generated by $R^{\underline \alpha} - R^{-\underline \alpha} $.

As the generators in $g_l$ form a representation of $E_{11}$ the above transformations for $g_0\in E_{11}$ can be written as 
$$
g_l\to g_0 g_lg_0^{-1}, \quad g_E\to g_0 g_E\quad {\rm and } \quad g_E\to g_E h
\eqno(1.3)$$
\par
The dynamics of the non-linear realisation is just an action, or set of equations of motion, that are invariant under the transformations of equation (1.2). We  now recall how  to construct the dynamics of the 
the $E_{11}\otimes_s l_1$ non-linear realisation using the  Cartan forms  which are given by 
$$
{\cal V}\equiv g^{-1} d g= {\cal V}_E+{\cal V}_l, 
\eqno(1.4)$$
where 
$$
{\cal V}_E=g_E^{-1}dg_E\equiv dz^\Pi G_{\Pi, \underline \alpha} R^{\underline \alpha}, \ {\rm and }\ 
{\cal V}_l= g_E^{-1}(g_l^{-1}dg_l) g_E= g_E^{-1} dz\cdot l g_E\equiv 
dz^\Pi E_\Pi{}^A l_A  
 \eqno(1.5)$$
Clearly ${\cal V}_E$ belongs to the $E_{11}$ algebra and it is  the Cartan form of $E_{11}$ while ${\cal V}_l$ is in the space of generators of the $l_1$ representation and one can recognise ${ E}_\Pi{}^A = (e^{A_{\underline \alpha}D^{\underline \alpha}})_\Pi{}^A$ as the vielbein on the generalised spacetime. 
\par
Both ${\cal V}_E$ and ${\cal V}_l$ are invariant under rigid transformations,  but  under the local $I_c(E_{11})$ transformations of equation (1.3) they change as 
$$ 
{\cal V}_E\to h^{-1}{\cal V}_E h + h^{-1} d h\quad {\rm and }\quad 
{\cal V}_l\to h^{-1}{\cal V}_l h 
\eqno(1.6)$$
 \par
To understand why the non-linear realisation leads to  equations of motion one just has to realise that the group element of equation (1.1) contains  the fields of the theory which depend on the generalised space-time. Hence writing down an invariant set of equations results in dynamical equations. 
In this way  one finds from the above procedure equations of motions for the fields which are either unique,  or almost  unique,  provided one specifies the number of derivatives involved. This understanding  goes back to the very earliest days of non-linear realisations, for example, to  the classic paper [18] which explained the procedure for the simplest kind of non-linear realisations.

%%%%%%%%%%%%%%%%%%%%%%%%%%%%%%%%%%%%%%%%%%%%%%%%%%%%%%%%%%%%%
\medskip
{\bf 2 The five dimensional theory}
\medskip
The theory in $D$ dimensions is found by deleting the  node labelled $D$ of the $E_{11}$ Dynkin diagram and decomposing the $E_{11}\otimes_s l_1$ algebra into representations of the resulting algebra. We choose in this paper to work in five dimensions and so we deleted node five  to find the algebra $GL(5)\otimes E_6$.
$$
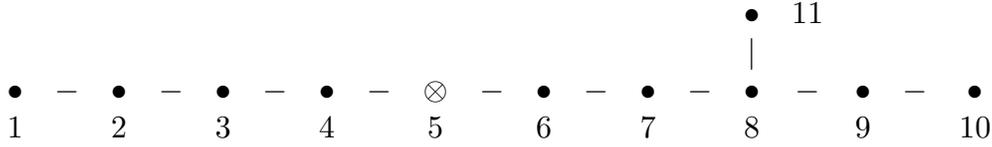
\matrix{
& & & & & & & & & & & & & & \bullet & 11 & & & \cr 
& & & & & & & & & & & & & & | & & & & \cr
\bullet & - & \bullet & - & \bullet & - & \bullet & - & \otimes & - & \bullet & - & \bullet & - & \bullet & - & \bullet & - & \bullet \cr
1 & & 2 & & 3 & & 4 & & 5 & & 6 & & 7 & & 8 & & 9 & & 10 \cr
}
$$
The Cartan involution invariant subalgebra of $GL(5)\otimes E_6$ is $I_c(GL(5)\otimes E_6)=SO(5)\otimes Usp(8)$. Since  the Cartan involution invariant subalgebra plays a central role in the construction of the dynamics using the Cartan forms we will decompose  the $E_{11}\otimes_s l_1$ algebra  into representations of $GL(5)\otimes Usp(8)$ rather than the algebra $Gl(5)\otimes E_6$. The further decomposition of representations of $GL(5)$ into those of $SO(5)$ being obvious. The decomposition of $E_{11}\otimes_s l_1$ into representations of $GL(5)\otimes E_6$ can be found  in the papers [7] and [17]. The level of an $E_{11}$ generator is just the number of up minus down $GL(5)$ indices.    
\par
The positive, including zero, level generators of the $E_{11}$ up to level 3 are 
$$
K^a{}_b,\ R^{\alpha_1 \alpha_2 }, R^{\alpha_1 \ldots \alpha_4 }, \  R^{a \alpha_1 \alpha_2 }, \  R^{a_1a_2}{}_{\alpha_1 \alpha_2 }, \ R^{a_1a_2a_3 \alpha_1 \alpha_2 }, R^{a_1a_2a_3 \alpha_1\ldots  \alpha_4 }, \ R^{a_1a_2,\,b} \ldots 
\eqno(2.1)$$
The indices $\alpha_1, \alpha_2,\ldots =1,\ldots ,8$ and we will use the Usp(8) invariant metric $\Omega_{\alpha_1 \alpha_2 }= \Omega_{[\alpha_1 \alpha_2 ]}$ to raise and lower indices as follows $T^\beta = \Omega ^{\beta \gamma }T_\gamma$,  $T_\alpha= \Omega _{\alpha\beta}T^\beta$  and so 
$\Omega _{\alpha\beta} \Omega ^{\beta \gamma}= \delta ^\gamma_\alpha$. The lower case Latin indexes correspond to 5-dimensional fundamental representation of $GL(5)$ ($a,\,b,\,c,\,... = 1,\,...,\,5$). The above generators also obey the relations  $R^{\alpha_1 \alpha_2 }= R^{(\alpha_1 \alpha_2) }$, $R^{a_1a_2a_3 \alpha_1 \alpha_2 }= R^{a_1a_2a_3 (\alpha_1 \alpha_2 )}$, $R^{[a_1a_2,\,b]} = 0$ while the indices on all the other generators are antisymmetric and $\Omega_{\alpha_1 \alpha_2 }$ traceless, for example 
$R^{\alpha_1 \ldots \alpha_4 } \Omega_{\alpha_1 \alpha_2 }=0$.  
The generators $R^{\alpha_1 \alpha_2 }$ are the generators of Usp(8) which  taken together with the generators $R^{\alpha_1 \ldots \alpha_4 }$  give   the algebra   $E_6$. The generators $R^{a \alpha_1 \alpha_2 }$ and $  R^{a_1a_2}{}_{\alpha_1 \alpha_2 }$ belong to the 27 and $\bar {27}$-dimensional representations of $E_6$ respectively. 
\par
The negative level generators are given by 
$$
R_{a\,\alpha_1\alpha_2},\quad R_{a_1a_2}{}^{\alpha_1\alpha_2},\quad R_{a_1a_2a_3 \alpha_1\alpha_2},\quad R_{a_1a_2a_3\,\alpha_1...\alpha_4},\quad R_{a_1a_2,\,b},\quad ...
\eqno(2.2)$$ 
The symmetries of their indices and the conditions they obey are  analogous to those given above for the positive level generators. 
\par
The vector, or $l_1$,   representation decomposes into representations of $Gl(5)\otimes Usp(8)$ to contain  
$$
l_A=\{ P_a,\quad Z^{\alpha_1\alpha_2},\quad Z^{a}{}_{\alpha_1\alpha_2},\quad Z^{a_1a_2 \alpha_1\alpha_2},\quad Z^{a_1a_2\,\alpha_1...\alpha_4},\quad Z^{ab},\quad ...\}
\eqno(2.3)$$
where $Z^{a_1a_2 \alpha_1\alpha_2}=Z^{a_1a_2\,\left(\alpha_1\alpha_2\right)}$ and the indices on all other generators are total antisymmetric and $\Omega_{\alpha_1 \alpha_2 }$ traceless except for the fourth generator $Z^{ab}$ which has no symmetries on its indices. The level for the vector representation  is the number of up minus down GL(5) indices plus one. The decomposition of the $E_{11}\otimes_s l_1$ algebra  into representations of $Gl(5)\otimes Usp(8)$ leads to many  equations even at low level and these commutators will be given in a longer paper [19].  
\par
 As explained in the introduction the  group element  of $E_{11}\otimes_s l_1$ can be written as  $g=g_l g_E$ where 
$$
g_l = \exp\{
x^a\,P_a + x_{\alpha_1\alpha_2} Z^{\alpha_1\alpha_2}+ x_{a}{}^{\alpha_1\alpha_2}Z^{a}{}_{\alpha_1\alpha_2}+ x_{a_1a_2 \alpha_1\alpha_2}Z^{a_1a_2 \alpha_1\alpha_2}
$$
$$
+ x_{a_1a_2 \alpha_1...\alpha_4}  Z^{a_1a_2\,\alpha_1...\alpha_4}+x_{ab} Z^{ab}+\ldots \}
\eqno(2.4)$$
$$
g_E =\ldots  g_3g_2g_1g_0
\eqno(2.5)$$
where 
$$
g_0= \exp{(h_{a}{}^{b}\,K^{a}{}_{b})}\,\exp{(\varphi_{\alpha_1\alpha_2}R^{\alpha_1 \alpha_2 }+ \varphi_{\alpha_1\ldots  \alpha_4 } R^{\alpha_1 \ldots \alpha_4 })}
\eqno(2.6)$$
$$
g_1= \exp{(A_{a}{}_{\alpha_1 \alpha_2 }R^{a}{}^{\alpha_1 \alpha_2 })},\quad 
g_2= \exp{( A_{a_1a_2}{}^{\alpha_1 \alpha_2 }R^{a_1a_2}{}_{\alpha_1 \alpha_2 })}, 
$$
$$
 g_3=\exp(A_{a_1a_2a_3 \alpha_1 \alpha_2 }R^{a_1a_2a_3 \alpha_1 \alpha_2 }+ A_{a_1a_2a_3 \alpha_1\ldots  \alpha_4 }R^{a_1a_2a_3 \alpha_1\ldots  \alpha_4 }+A_{a_1a_2, b} R^{a_1a_2,b})
\eqno(2.7)$$
In writing the group element $g_E$ we have used the local symmetry of the non-linear realisation of equation (1.2) to gauge away all terms that involve negative level generators in $g_E$. The group element $g_l$ is parameterised  by the quantities  
$$
x^a,\quad x_{\alpha_1\alpha_2},\quad x_{a}{}^{\alpha_1\alpha_2},\quad x_{a_1a_2 \alpha_1\alpha_2},\quad x_{a_1a_2\,\alpha_1...\alpha_4},\quad x_{ab},\ \ldots
\eqno(2.8)$$ 
which will be identified with the coordinates of the generalised space-time, In the group element $g_E$ we find the fields
$$
h_a{}^b,\ \varphi_{\alpha_1 \alpha_2 },\  \varphi_{\alpha_1 \ldots \alpha_4 }, \  A_{a \alpha_1 \alpha_2 }, \  A_{a_1a_2}{}^{\alpha_1 \alpha_2 }, \ A_{a_1a_2a_3 \alpha_1 \alpha_2 }, A_{a_1a_2a_3 \alpha_1\ldots  \alpha_4 }, \ A_{a_1a_2,\,b} , \ldots 
\eqno(2.9)$$
which depend on the coordinates of the generalised space-time.
\par
The $E_{11}$ Cartan forms for the five dimensional theory can be written in the form 
$$
{\cal V}_E= h_a{}^b K^a{}_b + G_{\alpha_1 \alpha_2 }R^{\alpha_1 \alpha_2 }+G_{\alpha_1 \ldots \alpha_4 } R^{\alpha_1 \ldots \alpha_4 }+ 
G_{a \alpha_1 \alpha_2 }R^{a \alpha_1 \alpha_2 }+ 
G_{a_1a_2}{}^{\alpha_1 \alpha_2 } R^{a_1a_2}{}_{\alpha_1 \alpha_2 }
$$
$$
+ G_{a_1a_2a_3 \alpha_1 \alpha_2 }R^{a_1a_2a_3 \alpha_1 \alpha_2 }+
G_{a_1a_2a_3 \alpha_1\ldots  \alpha_4 }R^{a_1a_2a_3 \alpha_1\ldots  \alpha_4 }+ G_{a_1a_2,\,b}R^{a_1a_2,\,b} + \ldots
\eqno(2.10)$$
The first index on the Cartan form is the one associated with the vector ($l_1$) representation  and it is not shown as we are using form notation, that is,  $G_{\underline \alpha} = dz^\Pi G_{\Pi , \underline \alpha}$. In what follows we will use the Cartan forms with their first index converted into a tangent index using the generalised vielbein, namely $E_A{}^\Pi G_{\Pi , \underline \alpha}=  G_{A , \underline \alpha}$. In what follows we will sometimes write the $E_{11}$ index as $\bullet$. 
\par
Using equation (1.6)  the variation of the Cartan forms under the Cartan invariant involution transformation  $I_c(E_{11})$ which involves the 
 generators at levels $\pm 1$  is given by 
$$
\delta\,{\cal V}_E = \left[S^{a\,\alpha_1\alpha_2}\,\Lambda_{a\,\alpha_1\alpha_2},{\cal V}_E\right] - S^{a\,\alpha_1\alpha_2}\,d\Lambda_{a\,\alpha_1\alpha_2}.
\eqno(2.11)$$ 
where 
$h=1- \Lambda _{a \alpha_1 \alpha_2 }S^{a \alpha_1 \alpha_2 }$ and  
$S^{a\,\alpha_1\alpha_2} = R^{a\,\alpha_1\alpha_2} - \eta^{ab}\,\Omega^{\alpha_1\beta_1}\,\Omega^{\alpha_2\beta_2}\,R_{b\,\beta_1\beta_2}$.
These  variations of the Cartan forms are straightforward to compute and are given by  
$$
\delta G_{ab} = 2\,G_{a\,\alpha_1\alpha_2}\,\Lambda_b{}^{\alpha_1\alpha_2} - {2\over 3}\,\eta_{ab}\,G_{c\,\alpha_1\alpha_2}\,\Lambda^{c\,\alpha_1\alpha_2} , \quad
 \delta G_{\left(\alpha_1\alpha_2\right)} = -\,4\,G_{a\,(\alpha_1\gamma}\,\Lambda^{a}{}_{\alpha_2)}{}^{\gamma} ,
\eqno(2.12)$$
$$
\delta G_{\alpha_1...\alpha_4} = 12\,G_{a\,[\alpha_1\alpha_2}\,\Lambda^a{}_{\alpha_3\alpha_4]} - 12\,\Omega_{[\alpha_1\alpha_2}\,G_{a\,\alpha_3\gamma}\,\Lambda^a{}_{\alpha_4]}{}^\gamma - \Omega_{[\alpha_1\alpha_2}\,\Omega_{\alpha_3\alpha_4]}\,G_{a\,\gamma_1\gamma_2}\,\Lambda^{a\,\gamma_1\gamma_2}.
\eqno(2.13)$$ 
$$
\delta G_{a\,\alpha_1\alpha_2} = -\,2\,G_{(ab)}\,\Lambda^b{}_{\alpha_1\alpha_2} - 2\,G_{\alpha_1\alpha_2\alpha_3\alpha_4}\,\Lambda_a{}^{\alpha_3\alpha_4} + 8\,G_{ab\,[\alpha_1\gamma}\,\Lambda^b{}_{\alpha_2]}{}^\gamma + \Omega_{\alpha_1\alpha_2}\,G_{ab\,\gamma_1\gamma_2}\,\Lambda^{b\,\gamma_1\gamma_2}.
\eqno(2.14)$$
$$
\delta G_{a_1a_2\,\alpha_1\alpha_2} = -\,4\,G_{[a_1\,[\alpha_1\gamma}\,\Lambda_{a_2]\,\alpha_2]}{}^\gamma - {1\over 2}\,\Omega_{\alpha_1\alpha_2}\,G_{[a_1\,\gamma_1\gamma_2}\,\Lambda_{a_2]}{}^{\gamma_1\gamma_2}
$$
$$
+\,6\,G_{a_1a_2b\,[\alpha_1\gamma}\,\Lambda^b{}_{\alpha_2]}{}^\gamma - 36\,G_{a_1a_2b\,\alpha_1\alpha_2\alpha_3\alpha_4}\,\Lambda^{b\,\alpha_3\alpha_4} - G_{a_1a_2,\,b}\,\Lambda^b{}_{\alpha_1\alpha_2}.
\eqno(2.15)$$
In deriving these equations we have taken into account the fact that the local transformations do not preserve the group element of equation (2.5-2.7). This requires some rather subtle steps which will be explained in [19]. 
\par
Using equation (1.6) we find that the first tangent index on the Cartan forms, that is the one associated with the vector representation, changes  under the local $I_c(E_{11})$ transformations as follows 
$$
\delta G_{a , \bullet } = -\Lambda_{a\,\alpha_1\alpha_2}G^{\alpha_1\alpha_2}{}_{, \bullet}\, \quad 
$$
$$
\delta G^{\alpha_1\alpha_2 }{}_{, \bullet} = 2\Lambda^{c\,\alpha_1\alpha_2} G_{c, \bullet} +4G^{c\gamma [\alpha_1}{}_{\bullet}\Lambda_c{}^{\alpha_2]}{}_\gamma - {1\over 2}\,\Omega_{\alpha_1\alpha_2}\,G^{a\gamma_1\gamma_2}{}_{,\bullet} \Lambda_{a\,\gamma_1\gamma_2},
$$
$$
\delta G_{a}{}_{\alpha_1\alpha_2}{}_{,\bullet } = 4 \Lambda_{a}{}_{\gamma[\alpha_1} G_{\alpha_2]}{}^{\gamma}{}_{\bullet}-{1\over 2}\,\Omega_{\alpha_1\alpha_2}\,G_{\gamma_1\gamma_2}{}_{,\bullet}\,\Lambda_a{}^{\gamma_1\gamma_2}.
\eqno(2.16)$$
Where $\bullet$ denotes the $E_{11}$ index that the Cartan forms carry. 
\par
The generalised vielbein can be easily computed, up to level one, from its definition in equation (1.5) to be given by 
$$
E_\Pi{}^A = \left(det\,e\right)^{-\,{1\over 2}}\,\left(\matrix{
e_\mu{}^a & -\,e_\mu{}^b\,A_{b\,\beta_1\beta_2} \cr
0 & f^{{\dot\alpha}_1{\dot\alpha}_2}{}_{\beta_1\beta_2} \cr
}\right),
\eqno(2.17)$$
where $e_\mu{}^a = \left(e^h\right)_\mu{}^a$ and $f^{{\dot\alpha}_1{\dot\alpha}_2}{}_{\beta_1\beta_2}$ is a function of the scalar fields which follows from its definition in this equation. 
\par
It is straightforward to compute the explicit form of the Cartan forms of equation (2.10) in terms of the fields that appear in the group elements of equation (2.4-2.7). One finds that the level zero Cartan forms are given by 
$$
G_a{}^b= (e^{-1})_a{}^\tau d e_\tau{}^b, \quad 
G^{\alpha_1\alpha_2}{}_{\beta_1\beta_2} - 2\,G^{[\alpha_1}{}_{[\beta_1}\,\delta_{\beta_2]}^{\alpha_2]}= \left(f^{-1}\right)^{\alpha_1\alpha_2}{}_{{\dot\gamma}_1{\dot\gamma}_2}\,df^{{\dot\gamma}_1{\dot\gamma}_2}{}_{\beta_1\beta_2}
\eqno(2.18)$$
The higher level contributions to the remaining  Cartan forms are given by 
$$
G_{\alpha_1 \ldots \alpha_4 } R^{\alpha_1 \ldots \alpha_4 }+ 
G_{a \alpha_1 \alpha_2 }R^{a \alpha_1 \alpha_2 }+ 
G_{a_1a_2}{}^{\alpha_1 \alpha_2 } R^{a_1a_2}{}_{\alpha_1 \alpha_2 }
+ G_{a_1a_2a_3 \alpha_1 \alpha_2 }R^{a_1a_2a_3 \alpha_1 \alpha_2 }
$$
$$+\ G_{a_1a_2a_3 \alpha_1\ldots  \alpha_4 }R^{a_1a_2a_3 \alpha_1\ldots  \alpha_4 }+ G_{a_1a_2,\,b}R^{a_1a_2,\,b}
$$
$$
= g_0^{-1} (\bar G_{\alpha_1 \ldots \alpha_4 } R^{\alpha_1 \ldots \alpha_4 }+ 
\bar G_{a \alpha_1 \alpha_2 }R^{a \alpha_1 \alpha_2 }+ 
\bar G_{a_1a_2}{}^{\alpha_1 \alpha_2 } R^{a_1a_2}{}_{\alpha_1 \alpha_2 }
+ \bar G_{a_1a_2a_3 \alpha_1 \alpha_2 }R^{a_1a_2a_3 \alpha_1 \alpha_2 }
$$
$$+\ \bar G_{a_1a_2a_3 \alpha_1\ldots  \alpha_4 }R^{a_1a_2a_3 \alpha_1\ldots  \alpha_4 }+ \bar G_{a_1a_2,\,b}R^{a_1a_2,\,b})g_0
\eqno(2.19)$$
 where 
$$
\bar G_{\mu\,\dot \alpha_1\dot \alpha_2} = dA_{\mu\,\dot \alpha_1\dot \alpha_2}, \quad
$$
$$
 \bar G_{\mu_1a_2\,\dot \alpha_1\dot \alpha_2} = dA_{\mu_1 a_2\,\dot \alpha_1\dot \alpha_2} - 2\,A_{[\mu_1\,[\dot \alpha_1\dot \gamma}\,dA_{\mu_2]\,\dot \alpha_2]}{}^{\dot \gamma}
- {1\over 4}\,\Omega_{\dot \alpha_1\dot \alpha_2}\,A_{[\mu_1\,\dot \gamma_1\dot \gamma_2}\,dA_{\mu_2]}{}^{\dot \gamma_1\dot \gamma_2},
$$
$$
\bar G_{\mu_1\mu_2\mu_3\,\dot \alpha_1...\dot \alpha_4} = \Big(dA_{\mu_1\mu_2\mu_3\,\dot \alpha_1...\dot \alpha_4} - A_{[\mu_1\,\dot \alpha_1\dot \alpha_2}\,dA_{\mu_2\mu_3]\,\dot \alpha_3\dot \alpha_4} + {2\over 3}\,A_{[\mu_1\,\dot \alpha_1\dot \alpha_2}\,A_{\mu_2\,\dot \alpha_3\dot \gamma}\,dA_{\mu_3]\,\dot \alpha_4}{}^{\dot \gamma}\Big)_{{{\rm proj}\  42}},
$$
$$
\bar G_{\mu_1\mu_2,\,\nu} = dA_{\mu_1\mu_2,\,\nu} - 2\,A_{\nu\,\dot \alpha_1\dot \alpha_2}\,dA_{\mu_1\mu_2}{}^{\dot \alpha_1\dot \alpha_2} + 2\,A_{[\nu\,\dot \alpha_1\dot \alpha_2}\,dA_{\mu_1\mu_2]}{}^{\dot \alpha_1\dot \alpha_2}
$$
$$
+\,{4\over 3}\,A_{\nu\,\dot \alpha_1\dot \alpha_2}\,A_{[\mu_1\,\dot \gamma}{}^{\dot \alpha_1}\,dA_{\mu_2]}{}^{\dot \gamma\dot \alpha_2} - {4\over 3}\,A_{[\nu\,\dot \alpha_1\dot \alpha_2}\,A_{\mu_1\,\dot \gamma}{}^{\dot \alpha_1}\,dA_{\mu_2]}{}^{\dot \gamma\dot \alpha_2},
$$
$$
\bar G_{\mu_1\mu_2\mu_3\,\left(\dot \alpha_1\dot \alpha_2\right)} = dA_{\mu_1\mu_2\mu_3\,\left(\dot \alpha_1\dot \alpha_2\right)} - 4\,A_{[\mu_1\,(\dot \alpha_1\dot \gamma}\,dA_{\mu_2\mu_3]\,\dot \alpha_2)}{}^{\dot \gamma}
$$
$$
+\,{4\over 3}\,A_{[\mu_1\,\dot \alpha_1\dot \gamma_1}\,A_{\mu_2\,\dot \alpha_2\dot \gamma_2}\,dA_{\mu_3]}{}^{\dot \gamma_1\dot \gamma_2} - {4\over 3}\,A_{[\mu_1\,(\dot \alpha_1\dot \gamma_1}\,A_{\mu_2}{}^{\dot \gamma_1}{}_{\dot \gamma_2}\,dA_{\mu_3]\,\dot \alpha_2)}{}^{\dot \gamma_2}
\eqno(2.20)$$
As  $g_0$ is a level zero  group element,  equation (2.19) holds separately at every level. Evaluating the effect of this group element one finds for the level one and two forms that 
$$
G_{a\,\alpha_1\alpha_2}= e_a{}^{\mu}\  \bar G_{\mu\,\dot  \delta_1 \dot \delta_2} f^{{\dot \delta}_1{\dot \delta}_2}{}_{\alpha_1\alpha_2}, \quad 
 G_{a_1a _2\,}{}^{\alpha_1\alpha_2}= e_{a_1}{}^{\mu_1}e_{a_2}{}^{\mu_2} \bar G_{\mu_1\mu_2\,}{}^{\dot \delta_1\dot \delta_2}
(f^{-1})^{\alpha_1\alpha_2}{}_{\dot \delta_1\dot \delta_2}
\eqno(2.21)$$
The net effect of the $g_0$ group element is to convert the world indices in both spacetime and internal space to be tangent indices. In the above the Usp(8) indices are denoted by $\alpha, \beta,\ldots $ if tangent and 
$\dot \alpha, \dot \beta,\ldots $ if world. In equation (2.20) no vielbeins are used to convert the indices on the fields, that is, the fields that appear, which  come from the group element, have  their indices  simply  replaced, for example  $a$ is replaced by  $\mu$. 
\medskip
{\bf 3 The equations of motion in five dimensions}
\medskip
We can now construct the equations of motion which, by definition, are those that are invariant under the symmetries of the non-linear realisations of equation (1.2), or equivalently,  the local variations of the Cartan forms of equation (2.12-2.15). We start from the viewpoint that the equations are first order in the generalised spacetime derivatives and so linear in the Cartan forms.  We begin by constructing the  equation of motion of the vector field and so this equation should contain the Cartan form $G_{a,b\alpha_1\alpha_2}$. We recall that, for example, the $a$ index was suppressed in the above variation as we use form notation. We will only consider  terms in the equations of motion   of the form $G_{\star , \bullet} $ where the vector index $\star$ can only take the level zero and one values, that is,  $a$ or $\alpha_1\alpha_2$ and the $E_{11}$ index $\bullet $ is below level five. We recall that level four contains,  the dual graviton and the dual scalar fields. Our aim in this paper is to find all terms  in the equations that contain the usual spacetime derivatives. However, as the terms which contain   derivatives with respect to the generalised level one coordinates can rotate, according to equation (2.16),  under the local symmetry into terms that have usual spacetime derivatives we must include these terms when this possibility arrises. This means we must include such terms  in the equation we are varying as they will lead to terms with the usual spacetime derivatives in the equation of motion that results from the variation.  
\par
One finds that the  equation which contains two SO(1,4) Lorentz indices and involves the vector field is given by 
$$
E^V_{a_1a_2\,\alpha_1\alpha_2} \equiv {\cal G}_{[a_1,\,a_2]\,\alpha_1\alpha_2}  + \theta\,G_{\alpha_1\alpha_2,\,[a_1a_2]}
\pm\,{1\over 2}\,\varepsilon_{a_1a_2}{}^{a_3a_4a_5}{\cal G}_{a_3,\,a_4a_5\,\alpha_1\alpha_2} =0,
\eqno(3.1)$$
where 
$$
{\cal G}_{[a_1,\,a_2]\,\alpha_1\alpha_2}\equiv  G_{[a_1,\,a_2]\,\alpha_1\alpha_2} + 2\,G_{[\alpha_1\gamma,\,a_1a_2\,\alpha_2]}{}^\gamma + {1\over 4}\,\Omega_{\alpha_1\alpha_2}\,G^{\gamma_1\gamma_2}{}_{, a_1a_2\,\gamma_1\gamma_2} 
\eqno(3.2)$$
and 
$$
{\cal G}_{a_3,\,a_4a_5\,\alpha_1\alpha_2}\equiv G_{a_3,\,a_4a_5\,\alpha_1\alpha_2} - G_{[\alpha_1\gamma,\,a_3a_4a_5\,\alpha_2]}{}^\gamma + 6\,G^{\alpha_3\alpha_4}{}_{, a_3a_4a_5\,\alpha_1...\alpha_4}
\eqno(3.3)$$
 We note the appearance in the vector equation of motion of derivatives with respect to the level one coordinates of the generalised spacetime. Those given in equations (3.2) and (3.3) are of the same form as those found from the $E_{11}\otimes_s l_1$ non-linear for the form fields in eleven [15] and four [16] dimensions. We also have such a term whose coefficient $\theta$ is  not determined at this stage of the calculation as it will lead to terms with the usual spacetime derivatives in the graviton equation. 
Apart from this parameter the equation is completely determined by the local symmetries at this stage of the calculation. 
\par
Under the local $I_c(E_{11})$ transformations the vector equation (3.1) transforms as 
$$
\delta E^V_{a_1a_2\,\alpha_1\alpha_2} = \mp\,2\,\varepsilon_{a_1a_2}{}^{a_3a_4a_5}\,E^V_{a_3a_4\,[\alpha_1\gamma}\,\Lambda_{a_5\,\alpha_2]}{}^\gamma \mp\,{1\over 4}\,\Omega_{\alpha_1\alpha_2}\,\varepsilon_{a_1a_2}{}^{a_3a_4a_5}\,E^V_{a_3a_4\,\gamma_1\gamma_2}\,\Lambda_{a_5}{}^{\gamma_1\gamma_2}
$$
$$
-\, 2\,E^S_{[a_1\,\alpha_1...\alpha_4}\,\Lambda_{a_2]}{}^{\alpha_3\alpha_4} \mp 2\,\varepsilon_{[a_1 |}{}^{b_1...b_4}\,\hat{E}^S_{b_1...b_4\,[\alpha_1 |\gamma}\,\Lambda_{ | a_2]\, |\alpha_2]}{}^\gamma - E^G_{a_1a_2,\,b}\,\Lambda^b{}_{\alpha_1\alpha_2}.
\eqno(3.4)$$
where $E^S_{a\,\alpha_1...\alpha_4}$ and $\hat{E}^S_{b_1...b_4\,\alpha_1\alpha_2}$ are  the scalar equations which is given by 
$$
E^S_{a\,\alpha_1...\alpha_4} \equiv  G_{a,\,\alpha_1...\alpha_4} \mp 6\,\varepsilon_a{}^{b_1...b_4}\,G_{b_1,\,b_2b_3b_4\,\alpha_1...\alpha_4}=0, 
\eqno(3.5)$$
$$
 \hat{E}^S_{a_1a_2a_3a_4 \alpha_1\alpha_2} \equiv G_{[a_1,\,a_2a_3a_4]\,\alpha_1\alpha_2}=0,
\eqno(3.6)$$
Equation (3.4) states that the variation of the vector equation gives back the vector equation $E^V_{a_1a_2\,\alpha_1\alpha_2}$, the scalar equations, $E^S_{a\,\alpha_1...\alpha_4}$ and $ \hat{E}^S_{a_1a_2a_3a_4 \alpha_1\alpha_2}$,  and an equation $E^G_{a_1a_2,\,b}$ corresponding to gravity. 
\par
Terms that contain  derivatives with respect to the higher level generalised coordinates are present in the scalar equations,  but are not needed at this stage of the calculation. We explain this point in more detail just below. To find them we must vary the scalar equation and this will be done in a future paper [19]. We now elucidate the appearance of such terms in more detail. Before  varying  a given  equation of motion , in addition to the terms that have the usual spacetime derivatives,  we must  add all terms with derivatives with respect to the level one generalised coordinates, that is,  all terms of the generic  form  
$$
k\times f \times G_{ \alpha _1 \alpha_2,\bullet }
\eqno(3.7)$$
 such that the term possess  the correct $SO(1,4)\otimes Usp(8)$ structure for a function $f$  of the Cartan forms. When we vary, using equation (2.16), such a term  we find the expression   $ -2k \Lambda ^c _{\alpha _1 \alpha_2}\times  G_{c ,\bullet }\times f$ and so a term of the generic form $G_{c ,\bullet }\times f$ in the equations of motion that we derive  from the variation of the original equation. However, when we vary this  new equation of motion we generate  terms, which  contain usual spacetime derivatives,    times the parameter  which must be cancelled. In this way the coefficients $k$ will be  determined. For the vector equation (3.1) all possible terms of the type of equation (3.7) have been added. However, the  terms in equation (3.2) and (3.3) are fixed in one step as they lead to terms in vector equation itself and  its variation gives back the same vector equation. On the other hand the term with parameter $\theta$ contributes to the gravity equation which we do not vary in this paper and so it is not determined at this stage of the calculation. 
\par
The gravity equation that emerges from the above calculation relates the usual spacetime derivative of the graviton to the usual spacetime derivative of the dual graviton by way of an alternating symbol. This was already found for $E_{11}\otimes l_1$ non-linear realisation in four [15]   and eleven [16] dimensions , however, as was also observed this equation holds modulo Lorentz transformations [15,16,20]. This  equation is both technically and conceptually very unfamiliar and as a result  it has not so far been understood how to process it. In this paper we will side step these difficulties. We will instead take  the space-time derivative of the vector equation (3.1) in such a way as to eliminate the dual vector field and so  find an equation of motion for the  vector equation which possess 
 two derivatives and so is of  the familiar form. 
 We will then proceed  by varying this equation under the local $I_c(E_{11})$ transformations to find the gravity equation which also has two derivatives and no dual graviton. 
\par
Using the form of the Cartan forms given in  equations (2.18-21) we find that after applying a space-time derivative to eliminate the dual vector field for the vector equation and the dual scalar fields for the scalar equation (3.5) 
 respectively that   these equations are explicitly given by 
$$
 e_{\mu_2}{}^a\,\partial_{\mu_1}\,\left[\left(\det{e}\right)^{{1\over 2}}\,G^{[\mu_1,\,\mu_2]}{}_{\beta_1\beta_2}\right] + G^{[b,\,a]}{}_{\delta_1\delta_2}\,G_{b}{}^{\delta_1\delta_2}{}_{\beta_1\beta_2} - 2\,G^{[b,\,a]}{}_{\delta[\beta_2}\,G_{b}{}^\delta{}_{\beta_1]}
$$
$$
\pm  \left(\det{e}\right)^{-1}\,\varepsilon^{ac_1...c_4}\,\left(G_{c_1,\,c_2\delta[\beta_1}\,G_{c_3,\,c_4}{}^{\delta}{}_{\beta_2]} + {1\over 8}\,\Omega_{\beta_1\beta_2}\,G_{c_1,\,c_2\delta_1\delta_2}\,G_{c_3,\,c_4}{}^{\delta_1\delta_2}\right) = 0,
\eqno(3.8)$$
and 
$$
 D_\mu\left[\left(\det{e}\right)^{{1\over 2}}  e_a{}^\mu\,G^a{}_{\alpha_1...\alpha_4}\right] \equiv \partial_\mu\left[\left(\det{e}\right)^{{1\over 2}}  e_a{}^\mu\,G^a{}_{\alpha_1...\alpha_4}\right] 
$$
$$+ 4\Big(G_{\mu,\,\delta[\alpha_1}\,G^{\mu,\,\delta}{}_{\alpha_2\alpha_3\alpha_4]}       
\Big)_{{{\rm proj}\  42}}
=-12\,\Big( G_{[c_1,\,c_2]{\dot\gamma}_1{\dot\gamma}_2}\,G^{[c_1,\,c_2]}{}_{{\dot\delta}_1{\dot\delta}_2} f^{{\dot\gamma}_1{\dot\gamma}_2}{}_{[\alpha_1\alpha_2}\,f^{{\dot\delta}_1{\dot\delta}_2}{}_{\alpha_3\alpha_4]}\Big)_{{{\rm proj}\  42}} 
\eqno(3.9)$$
In doing this we have dropped all terms that involve derivatives with respect to the higher level coordinates. We observe that these are precisely  the vector and scalar equations of five dimensional maximal supergravity. The curious factor of $\left(\det{e}\right)^{{1\over 2}}$ becomes a more familiar factor if one recalls that the  Cartan forms  with tangent indices contains the same factor by virtue of equation (2.17). 
\par
We now vary the vector equation of equation (3.8) under the local $I_c(E_{11})$ transformations of equation (2.12-2.15)  to find that we recover the scalar equation of motion (3.9) as well as the gravity equation which occurs as  the coefficient of $\Lambda_{\alpha_1\alpha_2}^b$. The results of this long and subtle calculation that involves several $Usp(8)$ identities  is the equation 
$$
(\det e) R_{ab}= 4G_{[a,c]}{}^{\delta_1 \delta_2} 
G_{[b,d]}{}_{\delta_1 \delta_2} \eta^{cd}
-{2\over 3}\eta_{ab} G_{[c,d]}{}^{\delta_1 \delta_2} 
G^{[c,d]}{}_{\delta_1 \delta_2} +k\  G_{a, \alpha_1\ldots \alpha_4}G_{b,}{}^{ \alpha_1\ldots \alpha_4}
\eqno(3.10)$$
where $R_{ab}$ is the Ricci tensor and $k$ is a constant. We  have also chosen the coefficient of the  term in the Ricci tensor, which is of  the form 
$R_{ab}\sim (e^{-1})_b{}^\mu\partial_\mu f_a$ for a suitable function $f_a$ of the vielbein. The ability to add these terms exploits the  mechanism explained around equation (3.7). The values of these coefficient will be determined once we vary the above equation under the $I_c(E_{11})$ transformations; we will report on this step elsewhere [19]. We recognise the left-hand side of equation (3.10) as the  energy momentum tensor for maximal five dimensional supergravity. 

%%%%%%%%%%%%%%%%%%%%%%%%%%%%%%%%%%%%%%%%%%%%%%%%%%%%%%%%%%%

\medskip 
{\bf 4 The eleven dimensional theory}
\medskip
In this section we will compute the $E_{11}\otimes_s l_1$ non-linear realisation  in eleven dimensions, so extending the results of [15]. The 
 theory in $D$ dimensions is found by deleting the  node labelled $D$ of the $E_{11}$ Dynkin diagram and decomposing the $E_{11}\otimes_s l_1$ algebra into representations of the resulting algebra. As such  we now deleted node eleven  to find the algebra $GL(11)$.
$$
\matrix{
& & & & & & & & & & & & & & \otimes & 11 & & & \cr 
& & & & & & & & & & & & & & | & & & & \cr
\bullet & - & \bullet & - & \bullet & - & \bullet & - & \bullet & - & \bullet & - & \bullet & - & \bullet & - & \bullet & - & \bullet \cr
1 & & 2 & & 3 & & 4 & & 5 & & 6 & & 7 & & 8 & & 9 & & 10 \cr
}
$$
\par
The decomposition of $E_{11}$ into representations of SL(11) has been given in many $E_{11}$ paper and it can also be found in the book  [27]. The level of an $E_{11}$ generator is the number of up minus down indices divided by three. 
The positive level generators are [1]
$$
K^a{}_b, \ R^{a_1a_2a_3}, \ R^{a_1a_2\dots a_6} \ {\rm and }\  R^{a_1a_2\ldots a_8,b}, \ldots 
\eqno(4.1)$$
where the generator
$R^{a_1a_2\ldots a_8,b}$ obeys the condition $R^{[a_1a_2\ldots a_8,b
]}=0$ and the indices $a,b,\ldots =1,2\ldots 11$.  
The  negative level generators  are  given by 
$$ 
R_{a_1a_2a_3}, \ R_{a_1a_2\dots a_6} , \ R_{a_1a_2\ldots a_8,b},\ldots
\eqno(4.2)$$
\par
The vector ($l_1$) representation decomposes into representations of $GL(11)$ as [3]
$$
P_a, Z^{ab}, \ Z^{a_1\ldots a_5}, \ Z^{a_1\ldots a_7,b},\  Z^{a_1\ldots a_8},\ 
Z^{b_1 b_2 b_3,a_1 ...a_8}, \ldots  
\eqno(4.3)$$
\par
The group element of $E_{11}\otimes_s l_1$ is of the form $g=g_l g_E$ where 
$$
g_E=  \ldots e^{ h_{a_1\ldots a_{8},b}
R^{a_1\ldots a_{8},b}} e^{ A_{a_1\ldots
a_6} R^{a_1\ldots a_6}}e^{ A_{a_1\ldots a_3} R^{a_1\ldots
a_3}} e^{h_a{}^b K^a{}_b}
\eqno(4.4)$$ 
and 
$$
g_l= e^{x^aP_a} e^{x_{ab}Z^{ab}} e^{x_{a_1\ldots a_5}Z^{a_1\ldots a_5}}\ldots = 
e^{z^A L_A} 
\eqno(4.5)$$
The fields and the generalised coordinates of the resulting theory can be read off from the group element. 
\par
The Cartan forms of $E_{11}$ can be written in the form 
$$
{\cal V} _E= 
G _{a}{}^b K^a{}_b+  G_{c_1\ldots c_3}
R^{c_1\ldots c_3} +G_{c_1\ldots c_6} R^{c_1\ldots c_6}+
G_{c_1\ldots c_8,b} R^{c_1\ldots c_8,b}+\ldots 
\eqno(4.6)$$
\par
They transform under the local $I_C(E_{11})$ transformation as dictated by equation (1.6). As the Cartan involution invariant subalgebra of SL(11) is SO(11) they transform under SO(11) for the  lowest level transformations. At the next level they transform under the group element 
$$
h=1-\Lambda_{a_1a_2a_3}S^{a_1a_2a_3}\ {\rm where}\ S^{a_1a_2a_3}= R^{a_1a_2a_3}- \eta^{a_1b_1} \eta^{a_2b_2}\ \eta^{a_1b_1} R_{b_1b_2b_3}
\eqno(4.7)$$
Under these latter transformations the Cartan forms of equation (4.7) transform as [15]
$$
\delta G^{ab}=18 \Lambda^{c_1c_2 b }G_{c_1c_2 }{}^{a}
-2 \delta ^{ab}  \Lambda^{c_1c_2 c_3}G_{c_1c_2 c_3},\ 
$$
$$
\delta G_{a_1a_2a_3}=-{5!\over 2} G_{b_1b_2b_3 a_1a_2a_3}
\Lambda^{b_1b_2 b_3}-6G^{c}{}_{[a_1 } \Lambda_{|c|a_2a_3]},\ 
$$
$$
\delta G_{a_1\ldots a_6}=2 \Lambda_{[ a_1a_2a_3}G_{a_4a_5a_6 ]}
-8.7.2 G_{b_1b_2b_3 [ a_1\ldots a_5,a_6]}\Lambda^{b_1b_2b_3}
+8.7.2 G_{b_1b_2[ a_1\ldots a_5a_6, b_3 ]}\Lambda^{b_1b_2b_3}
$$
$$
\delta G_{a_1\ldots a_8,b}=-3 G_{[ a_1\ldots a_6}\Lambda_{a_7a_8] b}
+3 G_{[ a_1\ldots a_6}\Lambda_{a_7a_8 b]}
\eqno(4.8)$$
\par
In the above the Cartan form were written as forms and so their first ($l_1$) index was suppressed. As before,  the Cartan form can be written as $G_{\star, \bullet}$ where the indices  $\star$  and $\bullet$ are associated with the vector and adjoint representations of $E_{11}$ respectively. Taking the former index to be a tangent index it transforms under the $I_c(E_{11})$ transformation of equation (4.7) as [15]
$$
\delta G_{a, \bullet}= -3G^{b_1b_2}{}_{,\bullet}\ 
\Lambda_{b_1b_2 a},
\quad \delta G^{a_1a_2}{}_{, \bullet}= 6\Lambda^{a_1a_2
b}  G_{b,}{}_{\bullet}
\eqno(4.9)$$
 \par
We now evaluate the $E_{11}$  Cartan form in terms of the field that parameterise the group element. We find that [15]
$$
{\cal V}_E = dz^\Pi G_{\Pi, \star} R^\star= G _{a}{}^b K^a{}_b +  G_{c_1\ldots c_3} R^{c_1\ldots c_3} +G_{c_1\ldots
c_6} R^{c_1\ldots c_6}+
G_{c_1\ldots c_8,b} R^{c_1\ldots c_8,b}+\ldots 
\eqno(4.10)$$
 Explicitly one finds that  [15]
$$
G _{a}{}^b=(e^{-1}d e)_a{}^b,\ \ G_{a_1\ldots a_3}= e_{a_1}{}^{\mu_1}\ldots e_{a_3}{}^{\mu_3}
dA_{\mu_1\ldots \mu_3}, 
$$
$$ 
G_{a_1\ldots a_6}=  e_{a_1}{}^{\mu_1}\ldots e_{a_6}{}^{\mu_6}(d A_{\mu_1\ldots \mu_6} 
- A_{[ \mu_1\ldots \mu_3}d A_{\mu_4\ldots \mu_6]})
$$
$$
G_{a_1\ldots a_8,b} =     e_{a_1}{}^{\mu_1}\ldots e_{a_8}{}^{\mu_8}e_{b}{}^{\nu}   (d h_{\mu_1\ldots \mu_8,\nu}
-A_{[\mu_1\ldots \mu_3}d A_{\mu_4\mu_5\mu_6} A_{\mu_7\mu_8 ]\nu}
+3 A_{[\mu_1\ldots \mu_6}d A_{\mu_7\mu_8 ]\nu}
$$
$$
+A_{[\mu_1\ldots \mu_3} d A_{\mu_4\mu_5\mu_6} A_{\mu_7\mu_8  \nu]}
-3 A_{[\mu_1\ldots \mu_6}d A_{\mu_7\mu_8 \nu]})
\eqno(4.11)$$
 where $e_\mu{}^a \equiv (e^h)_\mu{}^a$.  Equations (4.10) and (4.11) were given in reference [15]  and the slight differences to the equations of that reference are due to a different choice of parameterising the group element given in equation (4.4). 
 \par
The generalised vielbein  ${ E}_\Pi{}^A$, can be evaluated from its definition of equation (1.5) to be given as   a matrix  by [15]
$$
{ E}= (det e)^{-{1\over 2}}
\left(\matrix {e_\mu{}^a&-3 e_\mu{}^c A_{cb_1b_2}& 3 e_\mu{}^c A_{cb_1\ldots b_5}+{3\over 2} e_\mu{}^c A_{[b_1b_2b_3}A_{|c|b_4b_5]}\cr
0&(e^{-1})_{[b_1}{}^{\mu_1} (e^{-1})_{b_2]}{}^{\mu_2}&- A_{[b_1b_2b_3 }(e^{-1})_{b_4}{}^{\mu_1} (e^{-1})_{b_5 ]}{}^{\mu_2}  \cr
0&0& ( e^{-1})_{[b_1}{}^{\mu_1} \ldots (e^{-1})_{b_5]}{}^{\mu_5}\cr}\right)
\eqno(4.12)$$
\par
The non-linear realisation of $E_{11}\otimes_s l_1$ was computed at low  levels  in [15] where one  found that the three form and six form obey the equation 
$$
E_{a_1\ldots a_4}\equiv {\cal G}_{[a_1,a_2a_3a_4] }-{1\over 2.4!}\epsilon _{a_1a_2a_3a_4}{}^{b_1\ldots b_7} G_{[b_1,b_2\ldots b_7] }=0
\eqno(4.13)$$
where 
$${\cal G}_{a_1,a_2a_3a_4 }= G_{[a_1,a_2a_3a_4] }-{15\over 2}G^{b_1b_2}{}_{, b_1b_2 a_1\ldots a_4}
\eqno(4.14)$$
 On grounds of Lorentz invariance the only equation which is first order in the Cartan forms and has four Lorentz indices must be of the above generic form. It is far from obvious that it will also be  invariant under the higher level $I_c(E_{11})$ transformations of equation (4.7). However, the reader can easily verify that if one varies this equation under the transformations of equations (4.8) and (4.9), and one keeps  only the terms that involve the three form and six form, then the equation is invariant. One finds that [15]
 $$
\delta  E_{a_1\ldots a_4}= {1\over 4!} \epsilon _{a_1\ldots a_4 }{}^{b_1\ldots b_7} \Lambda_{b_1 b_2b_3} E_{b_4 \ldots b_7}+\ldots 
\eqno(4.15)$$
where $+\ldots$ denote gravity and dual gravity terms. 
 \par
 We now proceed as in the five dimensional case, rather than deduce the gravity equation by varying the above equation,  one can instead eliminate the dual gauge field, the six form, by taking a derivative in an appropriate way. One finds the equation 
 $$
\partial_{\nu}( (\det e)^{{1\over 2}} G^{[\nu,\mu_1\mu_2\mu_3]})+
{1\over 2.4!} (\det e)^{{-1}}\epsilon ^{\mu_1\mu_2\mu_3\tau_1\ldots\tau_8} G_{[\tau_1,\tau_2\tau_3\tau_4]} G_{[\tau_5,\tau_6\tau_7\tau_8] }=0
\eqno(4.16)$$
which is the familiar second order equation of motion for the three form. 
\par
To vary this equation under $I_c(E_{11})$ we rewrite it in terms of the  Cartan form of $E_{11}$  using the expressions of equation (4.11) to find that it is equivalent to the equation 
$$
E^{a_1a_2a_3}\equiv {1\over 2}  G_{b,d}{}^{d} G^{[b, a_1a_2a_3]}- 3G_{b,d}{}^{[a_1|} G^{[b, d |a_2a_3]]}
-G_{c,b}{}^{c} G^{[b, a_1a_2a_3]}
$$
$$
+ (\det e)^{{1\over 2}} e_b{}^\mu\partial_\mu G^{[b, a_1a_2a_3]}
+{1\over 2.4!}\epsilon ^{a_1a_2a_3b_1\ldots b_8} G_{[b_1,b_2 b_3 b_4]} G_{[b_5, b_6 b_7b_8] }=0
\eqno(4.17)$$
Vary the equation using the variations of the Cartan forms given in equation (4.8) and (4.9) and  reading off the coefficient of the parameter 
$\Lambda^{a_1a_2a_3}$ we find the equation 
$$
E_a{}^b\equiv (\det e) R_a{}^b- 48 G_{[a, c_1c_2c_3]}G^{[b, c_1c_2c_3]}+4\delta _a^b G_{[c_1, c_2c_3c_4]}G^{[c_1, c_2c_3c_4]}=0
\eqno(4.18)$$
which we recognise as the correct equation of motion for the graviton of the eleven dimensional supergravity theory. We have as in five dimension chosen the value of one  constant in the Ricci tensor. In varying to find this equation, as with other equations in this paper, we are adding terms to the vector equation of the form of equation (3.7). We will list these terms in a future paper [19].
\par
However, we will now vary this last gravity equation (4.18) under the transformations of equations (4.8) and (4.9) and fix this constant to be the value it was just chosen to be. Useful in this calculation is the variation of the spin connection which can be written in terms of the $E_{11}$ Cartan forms as 
$$
(\det e)^{{1\over 2}} \omega _{c, ab}= - G_{a, (bc)}+ G_{b, (ac)}+G_{c, [ab]}
\eqno(4.19)$$
one finds, at the linearised level,  that 
$$
\delta \omega _{c, ab}=-18 \Lambda ^{d_1d_2}{}_{ c} G_{[a,b]d_1d_2} -9 \Lambda ^{d_1d_2}{}_{ b} G_{a,c d_1d_2}+9 \Lambda ^{d_1d_2}{}_{ a} G_{b,c d_1d_2}
$$
$$
+2 \eta _{bc}  \Lambda ^{d_1d_2 d_3} G_{a, d_1d_2d_3}-2 \eta _{ac}  \Lambda ^{d_1d_2 d_3} G_{b, d_1d_2d_3}
\eqno(4.20)$$
plus terms that are a Lorentz transformation for  the linearised theory. 
\par
We will only carry out the variation at the linearised level where equation (4.18) takes the form 
$$
R_{ca}= \partial_c \omega_{b,}{}^{ab}-k\partial _b \omega _{c, }{}^{ab}
\eqno(4.21)$$
where in this equation we have indicated the constant $k$ which was so far undetermined.  Varying this equation,  under the $I_c(E_{11})$ transformation of equation (4.8), we  find that we recover the  correct result, that is, the linearised three form equation (4.16) only if one requires $k=1$ which 
is indeed the required result to get the Ricci tensor in equation (4.18). 
 In particular we find that 
$$
\delta R_{ca}= 36 \Lambda ^{d_1d_2 }{}_{c} \partial^b G_{[a,bd_1d_2]}
+36 \Lambda ^{d_1d_2 }{}_{a} \partial^b G_{[c,bd_1d_2]}+8 \eta_{ac} \Lambda ^{d_1d_2d_3 }\partial^b G_{[b,d_1d_2d_3]}
\eqno(4.22)$$
 Thus all the constants are fixed by the symmetries of the $E_{11}\otimes _s l_1$ non-linear realisation and the unique result is the bosonic  equations of motion of eleven dimensional supergravity. Further   details of the calculations in this  section will be given in reference [19].

%%%%%%%%%%%%%%%%%%%%%%%%%%%%%%%%%%%%%

\medskip
{\bf {5 Conclusion}}
\medskip
In this paper we have constructed the dynamics that follow from the non-linear realisation of $E_{11}\otimes_s l_1$  in five  dimensions and truncated the result to low levels, that is, we keep only the usual fields of supergravity and the usual coordinates of spacetime. We find the equations of motion of the bosonic sector of five dimensional maximal supergravity.  In deriving this result  we  have  chosen the value of two constants which are not determined by the calculation carried out in this paper, however, their  values will be determined once the calculation is extended [19]. We also found the dynamics of the $E_{11}\otimes_s l_1$   in eleven dimensions and, carrying out the same restrictions, we find precisely the bosonic equations of motion of eleven dimensional supergravity. In this last calculation there are no free constants as we took the variations under the symmetries of the non-linear realisation one step further. 
\par
The non-linear realisation provides a very direct path from the  algebra which defines the non-linear realisation    to the dynamics and so in the  case studied in this paper we have a direct path from the Dynkin diagram of $E_{11}$ to the equations of motion of  the five and eleven dimensional maximal supergravity theories. The assumptions are that we use the vector representation of $E_{11}$ to build the semi-direct product algebra and that we require to smallest number space time derivatives which leads to non-trivial dynamics. 
\par
As explained in the introduction the results in this paper strongly support the $E_{11}$ conjecture [1,2], which is that the low energy effective action of strings and branes is the non-linear realisation of $E_{11}\otimes_s l_1$. This is a unified theory that  includes in one theory  all the  maximal supergravity theories including the gauged supergravities. Thus we have a starting point from which to more systematically consider what is the underlying theory of strings and branes. We note that eleven dimensions does not play the preferred role it does in M theory as the theories in the different dimensions arise form different decompositions of $E_{11}$. Indeed,  the theories on finds in this way are completely equivalent, the coordinates and fields be rearranged from one theory to another corresponding to the different   decompositions of $E_{11}\otimes_s l_1$ being used  [5]. 
\par
The results found in this paper have been possible as a result of a better understanding of the consequences of fixing the local symmetry in the non-linear realisation and also how to technically process the field equations. A more detailed  account of the techniques will be given in [19].  
\par
There a quite a few avenues to explore. Perhaps the  first is to extend the calculation given in this paper to carry out the local $I_c(E_{11})$ variations of more of the equations of motion and also to carry out the variations at  a higher level,  especially by including terms that contain    derivatives with respect to  the higher level  coordinates. One can also return to the calculation of the dynamics formulated by taking  terms linear in the Cartan forms as this is likely to be the method most natural to include the much higher level fields. The next step in this  procedure would be to process the equation involving the dual graviton, using  the lessons learn in this paper,  and so recover the Einstein equation  from this view point. 
\par
The equations of motion that we have found are gauge  invariant even though we did not demand this was a symmetry from the outset. The gauge transformations appropriate to the non-linear realisation of $E_{11}\otimes_s l_1$  of were given in [21] and although these include the rigid $E_{11}$ transformations it is unclear if the gauge symmetries are automatically a symmetry of the higher level equations. It would be interesting to answer this question. 
\par
Although they were truncated out of most of our final equations,  the generalised coordinates beyond those of the usual spacetime play a crucial role in  the way the equations of motion were derived, indeed one could not derive these equations without them.  Given the  essential role they play,  the presence of   an $E_{11}$ symmetry and that the lowest level coordinate in the vector representation is that of spacetime there is  reason to believe in the existence of the higher level coordinates. We should think of these extra coordinates as leading to physical effects, indeed they required for the gauged supergravities [7]. It is very unlikely that our usual notion of spacetime 
survives in a fundamental theory of physics and in particular in the underlying theory of strings and branes. One can consider  the generalised coordinates used in this and previous $E_{11}$ papers as a kind of low energy effect theory of spacetime that represents the properties of spacetime before it is replaced by different degrees of freedom. This can be thought of as  analogous  to the usual  low energy effective actions which are built from fields that are replaced by different degrees of freedom in the underlying theory. The problem of how to eliminate all the higher level coordinates in the applications we are used to is a problem whose resolution demands a physical as well as a mathematical idea. 
\par
The  non-linear realisation of $E_{11}\otimes_s l_1$  contains an infinite number of fields whose role is only known for a small fraction of which.  However, $E_{11}$ has specific equations of motion including those for  the higher level fields and as such it is very predictive. It would be interesting to understand what role more of the higher level fields play. 
\par
Siegel theory [22,23], also called doubled field theory, is just a truncation of the non-linear realisation of $E_{11}\otimes_s l_1$   in ten dimensions  to the  lowest level zero [24]. The  extension of Siegel theory to include the fields of the massless R-R sector of the superstring was first given in reference [25] which  extended the  $E_{11}\otimes_s l_1$ non-linear realisation  theory to  level one. It must also be true that exceptional field theory [26],  which uses the  level zero and one coordinates [2] and ideas [1]  of the $E_{11}$ approach,  is essentially a truncation of the non-linear realisation of $E_{11}\otimes_s l_1$  to low levels. Indeed many of the equations of exceptional field theory were derived from the $E_{11}$ viewpoint in [17] including the underlying gauge transformations from which the exceptional theories are constructed. It is important to note that we have not required any a priori restriction on the way the fields depend on the generalised coordinates and one can think that the appearance of the section condition in the Siegel and exceptional field theories  is just a consequence of the brutal truncation required to obtain these theories.

\medskip
{\bf {Acknowledgements}}
\medskip
We wish to thank Nikolay Gromov for help with the derivation of the equations of motion from the non-linear realisation. We also wish to thank the SFTC for support from Consolidated grant number ST/J002798/1 and Alexander Tumanov wishes to thanks King's College  for the support provided by his  Graduate School International Research Studentship. 
\medskip
{\bf {References}}
\medskip
\item{[1]} P. West, {\it $E_{11}$ and M Theory}, Class. Quant.  
Grav.  {\bf 18}, (2001) 4443, hep-th/ 0104081. 
\item{[2]} P. West, {\it $E_{11}$, SL(32) and Central Charges},
Phys. Lett. {\bf B 575} (2003) 333-342,  hep-th/0307098. 
\item{[3]} See for example 
 B.~de Wit, H.~Samtleben and M.~Trigiante,
  {\it The maximal D = 5 supergravities},
  Nucl.\ Phys.\  B {\bf 716} (2005) 215, arXiv:hep-th/0412173; 
 B.~de Wit and H.~Samtleben,
  {\it Gauged maximal supergravities and hierarchies of nonabelian vector-tensor systems}, Fortsch.\ Phys.\  {\bf 53} (2005) 442, arXiv:hep-th/0501243 and references therein.
\item{[4]}I. Schnakenburg and  P. West, {\it Kac-Moody   
symmetries of
IIB supergravity}, Phys. Lett. {\bf B517} (2001) 421, hep-th/0107181.
\item{[5]} P. West, {\it The IIA, IIB and eleven dimensional theories 
and their common
$E_{11}$ origin}, Nucl. Phys. B693 (2004) 76-102, hep-th/0402140. 
\item{[6]}  F. ÊRiccioni and P. West, {\it
The $E_{11}$ origin of all maximal supergravities}, ÊJHEP {\bf 0707}
(2007) 063; ÊarXiv:0705.0752.
\item{[7]} ÊF. Riccioni and P. West, {\it E(11)-extended spacetime
and gauged supergravities},
JHEP {\bf 0802} (2008) 039, ÊarXiv:0712.1795.
\item{[8]} E. Bergshoeff, I. De Baetselier and  T. Nutma, {\it 
E(11) and the Embedding Tensor},  JHEP 0709 (2007) 047, arXiv:0705.1304. 
\item{[9]} A.~Kleinschmidt, I.~Schnakenburg and P.~West, {\it Very-extended Kac-Moody algebras and their interpretation at low  levels}, Class.\ Quant.\ Grav.\  {\bf 21} (2004) 2493 [arXiv:hep-th/0309198].; P.~West, {\it E(11), ten forms and supergravity},  JHEP {\bf 0603} (2006) 072,  [arXiv:hep-th/0511153]. 
\item{[10]} E.~A.~Bergshoeff, M.~de Roo, S.~F.~Kerstan and F.~Riccioni,
  {\it IIB supergravity revisited},JHEP {\bf 0508} (2005) 098
  [arXiv:hep-th/0506013]; E.~A.~Bergshoeff, M.~de Roo, S.~F.~Kerstan, T.~Ortin and F.~Riccioni, {\it IIA ten-forms and the gauge algebras of maximal supergravity theories},  JHEP {\bf 0607} (2006) 018
  [arXiv:hep-th/0602280]. 
\item{[11]}  A. Kleinschmidt and P. West, {\it  Representations of G+++
and the role of space-time},  JHEP 0402 (2004) 033,  hep-th/0312247.
\item{[12]} P. Cook and P. West, {\it Charge multiplets and masses
for E(11)}, ÊJHEP {\bf 11} (2008) 091, arXiv:0805.4451.
\item{[13]} P. West,  {\it $E_{11}$ origin of Brane charges and U-duality
multiplets}, JHEP 0408 (2004) 052, hep-th/0406150. 
\item{[14]} P. West, {\it Brane dynamics, central charges and
$E_{11}$}, JHEP 0503 (2005) 077, hep-th/0412336. 
\item {[15]} P. West, {\it Generalised Geometry, eleven dimensions
and $E_{11}$}, JHEP 1202 (2012) 018, arXiv:1111.1642.  
\item{[16]} P. West, {\it  E11, Generalised space-time and equations of motion in four dimensions}, JHEP 1212 (2012) 068, arXiv:1206.7045. 
\item{[17]} A. Tumanov and P. West, {\it E11 and exceptional field theory }, 
 arXiv:1507.08912.  
\item{[18]} S. Coleman, J. Wess and  B. Zumino, {\it Structure of phenomenological Lagrangians. 1.}, Phys.Rev. 163 (1967) 1727; 
C. Callan, S. Coleman, J. Wess and  B. Zumino,  {\it Structure of phenomenological Lagrangians. 2.}, Phys.Rev. 177 (1969) 2239. 
\item{[19]} A. Tumanov and P. West, to be published. 
\item{[20]} P. West, {\it Dual gravity and E11  },  arXiv:1411.0920. 
\item{[21]} P. West, {\it Generalised Space-time and Gauge Transformations},
JHEP 1408 (2014) 050,  arXiv:1403.6395.  
\item{[22]} W. Siegel, {\it Two vielbein formalism for string inspired axionic gravity},   Phys.Rev. D47 (1993) 5453,  hep-th/9302036; 
\item{[23]} W. Siegel,{\it Superspace duality in low-energy superstrings}, Phys.Rev. D48 (1993) 2826-2837, hep-th/9305073; 
{\it Manifest duality in low-energy superstrings},  
In *Berkeley 1993, Proceedings, Strings '93* 353,  hep-th/9308133. 
\item{[24]} P. West, {\it E11, generalised space-time and IIA string
theory}, 
 Phys.Lett.B696 (2011) 403-409,   arXiv:1009.2624.
\item{[25]} A. Rocen and P. West,  {\it E11, generalised space-time and
IIA string theory;  the R-R sector}, in {\it Strings, Gauge fields and the Geometry 
behind:The Legacy of Maximilian Kreuzer} edited by  Anton Rebhan, Ludmil Katzarkov,  Johanna Knapp, Radoslav Rashkov, Emanuel Scheid, World Scientific, 2013, arXiv:1012.2744.
\item{[26]}  O. Hohm and  H. Samtleben, {\it Exceptional Form of D=11 Supergravity}, Phys. Rev. Lett. 111 (2013)  231601, arXiv:1308.1673; 
 {\it Exceptional Field Theory I: $E_{6(6)}$ covariant Form of M-Theory and Type IIB},  Phys. Rev. D 89, (2014) 066016 , arXiv:1312.0614;  {\it Exceptional Field Theory II: E$_{7(7)}$} ,  arXiv:1312.0614; {\it   Exceptional Field Theory III: E$_{8(8)}$},  Phys. Rev. D 90, (2014) 066002, arXiv:1406.3348. 
\item{[27]} P. West, {\it Introduction to Strings and Branes}, Cambridge University Press, 2012. 

\end